# Arrangement of DOBAMBC molecules inside the capsule on change of the molecule's inclination on the border of the capsule investigated by the molecular dynamics method

*M.A. Korshunov*

L.V. Kirensky Institute of Physics Siberian Branch of RAS, 660036 Krasnoyarsk, Russia

e-mail: kors@iph.krasn.ru

**Abstract** The method of molecular dynamics is used to investigate the distribution of DOBAMBC molecules in a capsule with the fixed border layer. Change of an arrangement of molecules in smectic layers depending on an inclination of molecules on border is considered.

The arrangement of molecules of a liquid crystal affects light passage through the sample (for example, in displays). Therefore to know an arrangement of molecules in liquid crystals and possibility on it to influence it is represented an important practical problem. For this purpose to us numerical modelling is represented to the most perspective, using a method of molecular dynamics. Modelling allows to find an arrangement of molecules in structure of a liquid crystal and display of this arrangement at optical researches. More convenient for practical application, it appears to use film of liquid crystal composites which consist of a polymeric matrix in which cavities filled with molecules of a liquid crystal (a liquid crystal in capsules) there are. Such composites have a number of advantages in comparison with analogues on the basis of pure liquid crystals. It is high brightness, flexibility, cheaper manufacturing techniques, high reliability in operation. The big experimental material on capsulated to liquid crystals [1] is saved up. Among segnetoelectric liquid crystals to the most studied by various methods is DOBAMBC [2]. It is obviously important to study change of structure of a liquid crystal at change of boundary conditions.

The method of molecular dynamics is in detail described in a number of works [3-5]. The High light in this method is calculation of forces which are from potential. Therefore its choice is defining for the end result.

Development of methods of modelling of liquid crystals is described in work [6]. They were successfully used for modelling of liquid crystals with real molecules [7-9].

Molecules of a liquid crystal extended also can be bent that should be considered at calculations.

For this purpose the method of molecular mechanics [10] is used. The interaction potential is represented in a kind

$$U = U_s + U_b + U_t + U_{vdW} + U_e,$$

Where $U_s$ - potential energy of valency communications, $U_b$ - valency corners, $U_t$ - torsion corners, $U_{vdW}$ – interactions of Van-der-Vaalsa and $U_e$ - Coulomb forces.

For the description of intermolecular interaction the method atom-atom of potentials [11] was used. The potential of interaction $U_{vdW}$ has been used in shape

$$U_{vdW} = \sum_{i,j}(-\frac{A_{ij}}{r_{ij}^6} + B_{ij}exp^{(-c_{ij}r_{ij})}) .$$

In calculations $U_{vdW\ the}$ factors received by us earlier [12-14] were used.

Calculation of co-ordinates of atoms and their speeds in the course of interaction can be found, using algorithm of Verle in the high-speed form [15]. The step on time made 2fs.

Calculations were spent under the program written in language FORTRAN. At calculations technology CUDA [16-19] allowing was applied to use at the decision of problems graphic processors and to spend parallel calculations that has considerably raised speed of calculations on the COMPUTER.

The model capsulated a liquid crystal has been presented as follows. About a surfacelayers make a capsule cover, and interaction between these molecules and molecules in capsule volume was considered. To investigate, as the arrangement of molecules in a cover influences an arrangement of molecules in volume of a liquid crystal, calculations when cover molecules settled down Tangent were carried out. It changed force of interaction between a liquid crystal and a cover. Quantity of molecules in a capsule 2000.

Thus, as a result of calculations co-ordinates of atoms of molecules have been found at the given temperature that has allowed to find an arrangement of molecules in a capsule and to calculate order parametre <P2> and <P4> depending on temperature change (drawing 1).

The arrangement of molecules in the field of phase existence smectic $C^*$ thus in structure is received the coexistence of a number helicoidal fashions with the different period is marked. Because the arrangement of molecules non-uniform in separate areas of a liquid crystal can be found in each layer helicoidal fashions with the smaller period than for all crystal.

Besides change of an inclination of molecules in frontier area has allowed to note change of an arrangement of molecules of a liquid crystal from change of interaction with a cover.

In Capsule a liquid crystal reduction of temperature of transition from smectic with in smectic A (drawing 1) is observed. In comparison with experimental data on this crystal not being in a capsule [2].

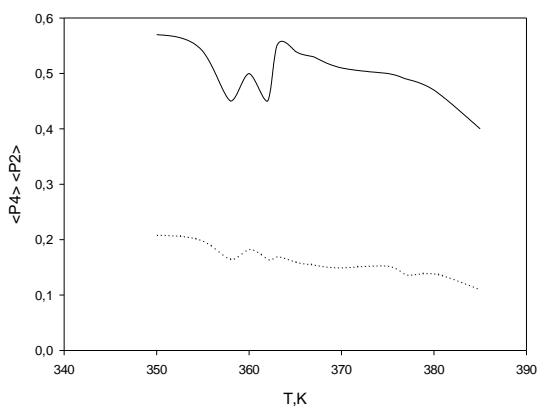

Drawing 1. Order parametre <P2> the top schedule, <P4> the bottom schedule depending on temperature for liquid crystal DOBAMBC in a spherical capsule.

In process of removal from border orderliness of molecules increases.

Change of an inclination of molecules in frontier area affects change of force of interaction between a liquid crystal and a capsule cover.

In work [2] the concept compression smectic layers and factor of compression $K_4$ is considered. This compression is defined by change of an inclination of molecules in a layer smectic at the given temperature. In drawing 2 change of an angle of slope of molecules in a layer is resulted at an arrangement of molecules of a cover in two positions radial (a) and axial (b). At a radial arrangement of molecules interaction between molecules of a cover and a liquid crystal is less than at an axial arrangement of molecules of a cover. It causes that change of orientation of molecules (compression of layers) begins at smaller temperature in a cover with a radial arrangement of molecules. Thus, interaction reduction between frontier molecules at their tangential arrangement and a liquid crystal in volume increases temperature of transition from smectic C* in smectic A.

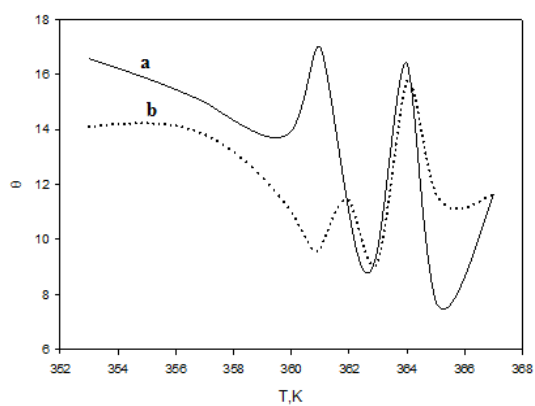

Drawing 2. Change in an inclination of molecules in a layer smectic at (a) a radial arrangement of molecules of a cover and axial (b) depending on temperature.

At experimental manufacturing capsulated a liquid crystal the sphere can have heterogeneity and defects. Therefore calculations of an arrangement of molecules DOBAMBC if on sphere there are defects have been carried out. It is found that in area with the increased interaction concentration of molecules of a liquid crystal increases, and in area with the lowered interaction decreases, but increases on its edges.

Thus, specially entering certain defects on a capsule surface it is possible to influence an arrangement of molecules in it.

Thus, numerical modelling of behaviour of molecules of liquid crystal DOBAMBC by a method of molecular dynamics is spent. Change of an arrangement of molecules in layers smectic from an inclination of molecules on border is considered. That changes force of interaction between molecules of a liquid crystal and frontier molecules. At a radial arrangement of molecules in a cover of a capsule change of an inclination of molecules in a layer smectic begins at smaller temperature, than at an axial arrangement of molecules of a cover. The behaviour of parametres of an order $<P_2>$ and $<P_4>$ depending on temperature is calculated. Presence on sphere of area with larger force of interaction increases about it concentration of molecules, and in area with the lowered interaction decreases, but increases on its edges. Calculations have shown that in nano to a capsule the temperature of transition from smectic phases C* in smectic phase A goes down. Thus, it is visible that capsule presence essentially affects an arrangement of molecules in volume of a liquid crystal.

**References**

1. *Dumrongrattana S., Huang C. C.* Polarization and tilt-angle measurements near the smectic-*A*–chiral–smectic-*C* transition of *P* - (*n*-decyloxybenzylidene) - P-AMINO - (2-methyl-butyl) cinnamate (DOBAMBC)//Phys. Rev. LETT.- V. 56.-No. 5.-C. 464-467.

2. *Blinov. L. M.* Structure and Properties of Liquid Crystals. Dordrecht Heidelberg London New York: Springer ScienceþBusiness Media B.V. 2011. 439c.

3. *Allen M.P., Tildesley D.J.* Computer Simulation of Liquids. Oxford: Clarendon Press, 1987.-385 p.

4. *Frenkel D., Smit B.* Understanding Molecular Simulation. From Algorithms to Applications. London: Academic Press, 1996. – 638 p.

5. *Rapaport D.C.* The art of molecular dynamics simulation. Cambridge: University Press, 1995.-549 p.

6. *Wilson M.R.* Progress in computer simulations of liquid crystals//Int. Rev. Phys. Chem. 2005 V. 24. P. 421-455.

7. *Berardi R., Zannoni C., Lintuvuori J.S. and. Wilson M.R.* A soft-core Gay–Berne model for the simulation of liquid crystals by Hamiltonian replica exchange//J. Chem. Phys. 2009. V.131. P. 174107-1-174107-6

8. *Pelaez J., Wilson M.* Molecular orientational and dipolar correlation in the liquid crystal

mixture E7: a molecular dynamics simulation study at a fully atomistic level//Phys. Chem. Chem. Phys. 2007 V. 9. P. 2968–2975.

9. *Picken S.J., van Gunsteren W.F., van Duijnen P.Th. and de Jeu W.H.* A molecular dynamics study of the nematic phase of 4-n-pentyl-4 '-cyanobiphenyl//Liquid Crystals. 1989 V. **6. P.** 357-371.

10. *Clark* T. Computer chemistry. M: the World, 1990.-380 c.

*11. Kitaigorodsky A.I.molecular* crystals. M: the Science, 1971. – 424 c.

12. *Shabanov V. F, Spiridonov V. P, Kites of M. And.* Polarising researches of a spectrum of combinational dispersion of small frequencies paradibrom - and paradichlorbenzol //Zhurn. prikl. A spectrum. 1976. V. 25. № 4. With. 698-701.

13 *Korshunov. M. A. and Shabanov V. F.* Size Effects on Dynamics of a P_Dibromobenzene Lattice. / Nanotechnologies in Russia, 2010, Vol. 5, Nos. 1–2, pp. 73–77.

14. *Korshunov M. A.* Low-Frequency Raman Spectra of Paradichlorobenzene Thin Films. / Optics and Spectroscopy, 2009, Vol. 106, No. 3, pp. 347–349.

15. *Swope W.C., Andersen H.C., Berens P.H. end Wilson K.R. A* Computer simulation method for the calculation of equilibrium constants for the formation of physical clusters of molecules: application to small water clusters.//J. Chem. Phys. 1982 V. **76. P.** 637 649.

16. v*an Meel J.A., Arnold A., Frenkel D., Zwart S.F.P., and Belleman R.G.* Harvesting graphics power for MD simulations.//Molecular Simulation., 2008. Vol. 34, No. 3. P. 259–266.

17. *Swope W.C., Andersen H.C., Berens P.H., Wilson K.R. A* Computer simulation method for the calculation of equilibrium constants for the formation of physical clusters of molecules: application to small water clusters//J. Chem. PHYS. - V. 76. **P.** 637 649.

18. *Meel J.A., Arnold A., Frenkel D., Zwart P.S.F., Belleman R.G.* Harvesting graphics power for MD simulations//Molecular Simulation.-2008.-V. 34.-No. 3. P. 259-266.

19. S*unarso A., Tsuji T., Chono S.*GPU-accelerated molecular dynamics simulation for study of liquid crystalline flows//Journal of Computational Physics.-2010.-V. 229.-P. 5486–5497.